\documentstyle[editedvolume,psfig]{crckapb} 

\begin{opening}
\title{Circumstellar Molecular Spectra towards Evolved Stars}

\author{Eric J. Bakker}
\institute{Department of Astronomy and W.J. McDonald Observatory,
University of Texas, Austin, TX 78712, USA}
\end{opening}

\runningtitle{Circumstellar Molecular Spectra}
\begin{document}

\begin{abstract}
In this paper we discuss the relevance 
of, and possible scientific gains which can be acquired from
studying circumstellar molecular spectra toward
evolved stars. Where can we expect circumstellar molecular spectra,
why would we want to study these spectra, which molecules 
might be present, and what can we learn from these studies?
We present an overview of reported detections, and discuss some of the
results.
\end{abstract}

\section{Introduction}

Observational studies of the circumstellar environment (CSE) of evolved stars
mainly concentrate on infrared emission from dust (peaking around 60 $\mu$m)
and radio line emission of various molecular species (e.g. CO, HCN, CN).
Both tracers are very powerful in studying a large sample of stars,
and the overall characteristics of the CSE, but, since
they are in emission they lack positional information about the line
forming region.

Other tracers of the CSE are infrared emission 
features (Polycyclic Aromatic
Hydrocarbons, PAH; Unidentified Infrared Bands, UIR; circumstellar reddening).
The physics of these latter tracers are not well understood and with
our current knowledge it is not possible to extract detailed physical 
information from
these tracers. 

A different category of tracer is molecular  spectra. Absorption
spectra only probe a well defined pencil beam toward the continuum source,
and are very well suited for detailed studies of 
the CSE. Emission spectra are rare and their occurrence is almost always
connected with some kind of violent mechanism (e.g. shocks).
Fortunately, the molecules for which circumstellar spectra have been
detected are rather simple  with at most three atoms (Table 1).
The physics of these 
molecules is sufficiently well understood that 
they can be used to extract physical information about the line 
forming region (Sect. 1.4).
An additional advantage of these molecules is that their 
electronic spectra fall in the ultraviolet and optical part 
of the spectrum and acquiring these spectra is relatively easy.

In this paper we discuss circumstellar molecular spectra with
a strong emphasis on absorption line studies.
We will ask ourself four question:
where are they, why bother, which species, and what can we learn?

\subsection{Where can we expect circumstellar spectra?}

In the broadest sense we could argue that circumstellar absorption spectra
are expected to be present in all cases where there is 
circumstellar matter between the observer and the continuum source.
The continuum source is mostly the star.
We therefore can expect molecular spectra
towards  RGB, AGB, post-AGB stars (Bakker et al. 1996/1997, 
Bakker \& Lambert, from now on Paper I, II, III), 
supernovae (Dalgarno 1998), Young Stellar Object
(Mitchell et al. 1991), and interacting binary systems.
Whether or not these absorption lines will 
be detected depends on the signal-to-noise ratio of the spectrum
in combination with the column density of the observable species.
This suggests that with the rise of larger telescopes and
more sensitive detectors, we expect to find many more objects
exhibiting circumstellar spectra.

\subsection{Why study molecular spectra as opposed to radio emission lines?}

Studies of radio line emission are faced with the problem that
for a spatially unresolved source, the emission profile 
is the integrated emission of the CSE. The intensity
at a given frequency, does not correspond with one location in 
the shell, but with a ring of material for which the projected
velocity in the radial direction corresponds to that  frequency.
If the shell is not spherically symmetric this problem is even
more complicated.
Both these problems do not enter into absorption line studies:
velocities are not projected and asymmetry
is not important. Only inhomogeneities along the line of sight
could be important.

\subsection{Which molecular species might be present?}

In order to give the final answer to this question, one has to know
all allowed transitions of all molecules possibly present. 
This is beyond the scope of our work. 
Instead we look at studies concerning interstellar, photospheric, 
and cometary molecules. Absorption lines  of molecules detected
in interstellar diffuse clouds are among others: H$_{2}$, HD, CH,
CH$^{+}$, C$_{2}$, OH, CO, NH, CN, CS(mm), and HCl 
(Van Dishoeck 1997).
Molecules detected in the photosphere of cool stars are among
others: C$_{2}$, CN, CH, HCN, and C$_{2}$H$_{2}$ 
(J{\o}rgensen 1997).
Molecules detected in cometary spectra are among others:
CO, C$_{2}$, CS, NH$_{3}$ and many highly reactive species
(Weaver 1997, Bockelee-Morvan 1997). An alternative approach
is to compute chemical models of the CSE 
(Cherchneff et al. 1993) and predict column densities. 

The question now arises which of these molecules are likely to
be present at high enough abundance to be detectable.
Cometary lines are due to evaporation and
successive photo-dissociation. This time scale is much shorter than
the corresponding time scale in circumstellar shells and 
cometary molecules (especially the reactive ones) might not be  abundant
in circumstellar shells. Photospheric molecules exist
at a much higher temperature than circumstellar molecules. 
The chemical equilibrium is different in the two environments
and therefore different molecules might be present. Interstellar
molecules experience roughly the same conditions (density, temperature,
and interstellar radiation field) as circumstellar molecules. In order
of probability of the presence of circumstellar molecules (with absorption
spectra) we propose interstellar, followed by photospheric and cometary species.
As we will demonstrated in this paper, circumstellar absorption spectra of
CH$^{+}$, C$_{2}$, CN, CO, C$_{3}$, SiC$_{2}$ and emission of
CH$^{+}$, C$_{2}$,     CO, C$_{3}$, AlO,  VO, and 
ZrO have been detected (Table 1). There remains
a large	 amount of work to be done to check for the presence of the 
other molecules. In particular the detection 
(at high-resolution, $\lambda/\Delta \lambda \geq 30,000$)
of circumstellar absorption of H$_{2}$ and CO is of fundamental importance.

\subsection{What can we learn from molecular spectra?}

{\sl Identification:}
      A symmetric molecule like C$_{2}$ does not have
      transitions in the radio. Electronic 
      bands allow to study these molecule.
      The interstellar 
      abundance of CH$^{+}$ is still not well understood, studying
      circumstellar CH$^{+}$ spectra might help to improve
      our understanding of the formation process. 
      The presence of the unidentified 21 $\mu$m feature
      is strongly correlated with the presence of C$_{2}$ and CN absorption,
      which suggests that the 21 $\mu$m feature is due to a carbon bearing
      molecule. \newline
{\sl Expansion velocity:}
      Very accurate expansion velocities can be determined.
      The correlation of expansion velocity with the photospheric abundance
      pattern (mainly carbon abundance) might give information about the
      mass-loss mechanism.
      Different molecules probe different layers. The expansion velocity
      can be determined throughout the shell and possibly a slow
      change of expansion velocity can be detected as a result
      of the star's evolution.  The temperature
      and chemical content of the shell can be studied as a function of radius.
      \newline
{\sl Non-LTE effects:}
      This is important for radio emission lines
      since in the radio only few levels of a molecule are probed and the 
      total number of molecules is obtained by extrapolating to higher levels 
      given a certain LTE temperature. We found that C$_{2}$ is pumped
      by the stellar radiation field (supra-thermal), while CN
      is sub-thermal (Paper I, II). \newline
{\sl Isotope ratios:}
      Nucleo-synthesis of AGB stars and the third dredge-up.
      Fig.~1 shows our first detection of $^{13}$CN Violet System (0,0)
      in HD56126 (Paper III). We found $^{12}$CN/$^{13}$CN=23$\pm$1
      and that this ratio is affected by the isotopic exchange reaction 
      and puts only a lower limit on the $^{12}$C/$^{13}$C ratio.
      \newline
{\sl Inhomogeneities (clumps):} 
      A possible change with time might be 
      detected of the molecular bands due to inhomogeneities in the line 
      of sight. Some of the stars pulsate. A time delay between the pulsation
      of the star and the change in temperature of the molecule, would yield
      the radius at which that molecule is present.

\begin{table}[htb]
\begin{center}
\caption{Circumstellar molecular lines in the optical and ultraviolet 
spectra of evolved stars.}
\begin{tabular}{llllllll} % 8 columns
\hline
\hline
Post-AGB stars:   &C$_{2}$&CN &21$\mu$m&[s/Fe]&C/O    &Remarks&References   \\
\cline{1-6}
IRAS02229+6208    &-      &-  &        &      &       &       &1,1,.,.,.    \\
IRAS04296+3429    &-      &-  &+       &+     &       &       &2,2,4,8,.    \\
IRAS05113+1347    &-      &-  &+       &      &       &       &2,2,5,.,.    \\
IRAS05341+0852    &-      &-  &+       &+     &+      &       &2,2,5,9,9    \\
HD56126           &-      &-  &+       &+     &+      &       &2,2,4,10,10  \\
IRAS07431+1115    &-      &-  &        &      &       &       &1,1,.,.,.    \\
IRAS08005-2356    &-      &-  &        &      &       &       &2,2,.,.,.    \\
IRAS14429-4539    &+?     &   &        &      &       &       &3,.,.,.,.    \\
HD187885          &nd     &nd &+       &+     &+      &       &24,24,6,11,11\\
IRAS20000+3239    &-      &-  &+       &      &       &       &2,2,5,.,.    \\
AFGL2688          &-+     &-  &+       &      &       &SiC$_{2}$ detected&2,2,.,.,.\\
IRAS22223+4327    &-      &-  &+       &+     &+      &       &2,2,5,8,8    \\
HD235858          &-      &-  &+       &+     &+      &       &2,2,4,12,12  \\
IRAS22574+6609    &-?     &-? &+       &      &       &       &24,24,7,.,.  \\
IRAS23304+6147    &-      &-  &+       &      &       &       &2,2,4,.,.,   \\
AGB stars:        &C$_{2}$&CN &21$\mu$m&[s/Fe]&C/O    &       &             \\
\cline{1-6}
IRC+10216         &-?     &-  &-       &+     &+      &       &2,2,.,22,23  \\
Post-AGB binaries:&C$_{2}$&CN &CH$^{+}$&CO    &C$_{3}$&       &             \\
\cline{1-6}
HD44179           &nd     &+  &+       &-+    &+      &       &2,19,2,20,21 \\
HD46703           &nd?    &nd?&nd?     &      &       &       &24,24,24,.,.,\\
HD52961           &nd     &nd&nd       &      &       &       &2,2,2,.,.    \\
HR4049            &nd     &nd &nd      &      &       &       &2,2,2,.,.    \\
HD213985          &nd     &nd &-       &      &       &       &2,2,2,.,.    \\
BD+39$^{o}$4926   &nd     &nd &nd      &      &       &       &2,2,2,.,.    \\
R CrB stars:      &C$_{2}$&CN &        &      &       &       &             \\
\cline{1-3}
V854 CEN          &+      &+  &        &      &       &       &16,17        \\
R CrB             &+      &   &        &      &       &       &18,.         \\
Anomalous objects:&AlO    &ScO&TiO     &VO    &       &       &             \\
\cline{1-5}
VY CMa            &       &+  &+       &      &       &disk ? &.,13,14,.    \\
U Equ             &+      &   &+       &+     &       &disk ? &15,.,15,15   \\
\hline  
\hline
\end{tabular} \newline
-: absorption; + emission or enhanced;
open slots if no information available \newline
1: Hrivnak 1998;
2: Bakker et al. 1997;
3: Hu et al. 1994;
4: Kwok et al. 1989;
5: Kwok et al. 1995;
6: Justtanont et al. 1996;
7: Hrivnak \& Kwok 1991;
8: Decin et al. 1998;
9: Reddy et al. 1997;
10: Klochkova 1995;
11: Van Winckel et al. 1996;
12: Za{\v c}s et al. 1995;
13: Wallerstein 1986;
14: Phillips \& Davis 1987;
15: Barnbaum et al. 1996;
16: Rao \& Lambert 1993;
17: Withney et al. 1992;
18: Lambert et al. 1990;
19: Sarre 1996;
20: Glinski et al. 1997;
21: Glinski \& Nuth 1995; 
22: Utsumi 1970; 
23: Lambert et al. 1986;
24: this paper
\end{center}
\end{table}

\begin{figure*} % figure 1
\centerline{\hbox{\psfig{figure=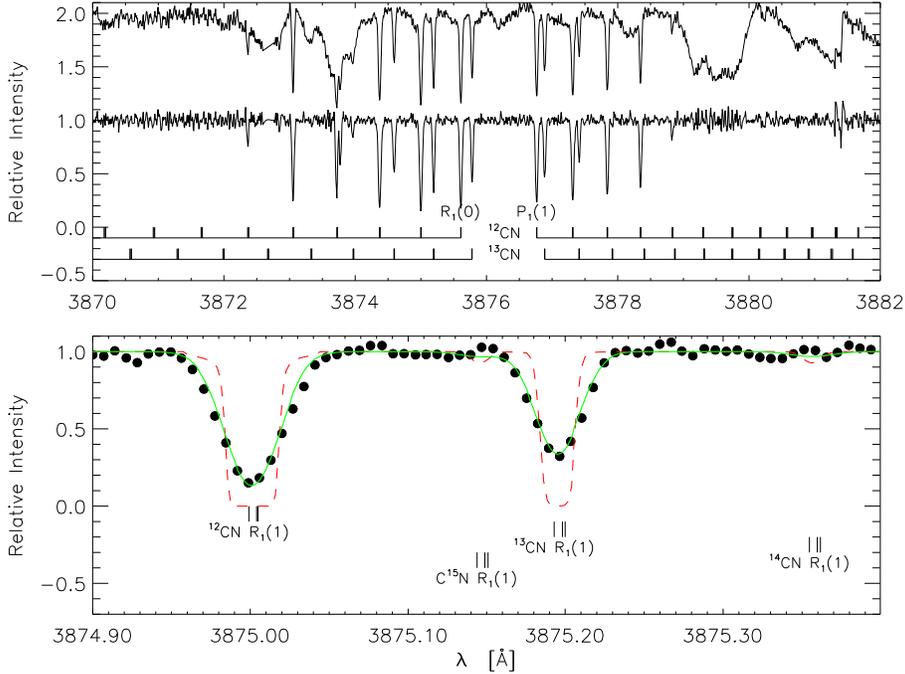,width=\textwidth}}}
\caption{CN Violet System (0,0) band towards HD~56126. 
The upper panel shows the observed spectrum and
a  rectified spectrum corrected
for the underlying photospheric features.
The rectified spectrum contains only circumstellar lines.
The lower panel shows on an expanded wavelength scale
the strongest line (the R$_{1}$, R$_{2}$, and $^{\rm R}Q_{21}$ blend 
for $N''=1$),
and demonstrates that we have not detected $^{14}$CN nor C$^{15}$N.
The dashed spectrum is a synthetic spectrum computed for $b=0.51$ km s$^{-1}$,
$T_{\rm rot}=12.6$ K for CN, $T_{\rm rot}=8.0$  for the CN isotope,
and the isotope or lower limit isotope ratios as determined
in this work. The solid spectrum is the synthetic spectrum 
convolved to a spectral resolution of $R=140,000$.
(Figure from Paper III).}
\end{figure*}

\section{Discussion}

Table~1 gives a summary of evolved objects for which circumstellar
molecular spectra have been detected. Only those objects and molecules
are listed for which there is ample evidence that they are indeed 
circumstellar. This could be based on the radial velocity, 
or on the temperature derived from the molecular band. Several groups
have conducted low-resolution studies of CO
(Hrivnak et al. 1994, Oudmaijer et al. 1995), 
C$_{2}$, CN, and C$_{3}$ (Hrivnak 1995). These
studies have not been included in this table because they do 
not make a distinction between photospheric and circumstellar origin.
The column headed by 21 $\mu$m, refers to the
presence of the unidentified 21 $\mu$m feature, [s/Fe] refers to the 
photospheric abundance of s-process elements relative to Fe, relative to
the solar ratio. C/O is the photospheric carbon over oxygen ratio.

Based on Table 1, we suggest three criteria which can be used
to identify a post-AGB star: A. the presence of C$_{2}$ and CN
absorption spectra; B. the presence of the 21 $\mu$m emission feature,
and C. an enhancement of s-process elements and carbon.

Of the large range of circumstellar molecules that could exhibit
molecular spectra in the ultraviolet, optical, or infrared 
(see Sect. 1.3), only a handful have been detected (Table 1).
The detection of more molecules,
but also the increase of objects exhibiting circumstellar
spectra, would give valuable information on circumstellar chemistry
and stellar evolution. It should be noted that in this paper we
have emphasized on molecular absorption spectra
and to a lesser extend molecular emission spectra. Besides molecules,
there are also atoms present which will exhibit absorption lines
from their ground level (Morton 1991).

In order to facilitate the search for circumstellar molecular bands,
extensive line list of the relevant transitions presented in Paper
I, II, and II, are (or will be) available at CDS, or can be obtained
by contacting one the author.

{\sl Acknowledgments}
The author acknowledges the National Science Foundation (Grant No.
AST-9315124) and the Robert A. Welch Foundation of Houston, Texas.
This research has made use of the Simbad database, operated at
CDS, Strasbourg, France, and the ADS service.

\end{document}